\RequirePackage{fix-cm}
\documentclass[twocolumn,epjc3]{svjour3}  

\smartqed  
\usepackage{color}
\usepackage{xcolor}
\definecolor{bred}{rgb}{0.8, 0.0, 0.0}
\definecolor{pblue}{rgb}{0.2, 0.2, 0.6}
\definecolor{ao}{rgb}{0.0, 0.5, 0.0}
\definecolor{carmine}{rgb}{0.59, 0.0, 0.09}

\newcommand{\CEvNS}{CE$\nu$NS}
\newcommand{\conusplus}{CONUS\texttt{+}}

\newcommand{\Kshell}{K\nobreakdash-shell}
\newcommand{\Lshell}{L\nobreakdash-shell}
\newcommand{\Mshell}{M\nobreakdash-shell}

\usepackage{upgreek}
\usepackage{subcaption}
\usepackage{url}
\usepackage{gensymb}
\usepackage{tabularx}
\usepackage{graphicx}
\usepackage{amsmath}
\usepackage{amssymb}
\usepackage{hyperref}
\usepackage{cite}
\usepackage{arydshln}
\usepackage{multirow}

\usepackage{hyperref}
\hypersetup{
    colorlinks = true,
    linkbordercolor = {white},
    linkcolor = bred,
    citecolor = pblue,
    urlcolor = pblue
}

\usepackage[switch]{lineno} 

\begin{document}\sloppy

\title{Sub-keV energy calibration of \conusplus\ via $^{71}$Ge M-shell neutron activation}
\titlerunning{Sub-keV energy calibration of \conusplus\ via $^{71}$Ge M-shell neutron activation}


\author{E.~S\'{a}nchez Garc\'{i}a\thanksref{e1, MPIK}, Y.~Shi\thanksref{e2, MPIK}, N.~Ackermann\thanksref{MPIK}, H.~Bonet\thanksref{MPIK}, C.~Buck\thanksref{MPIK}, J.~Hakenm\"{u}ller\thanksref{MPIK, Wien},  G.~Heusser\thanksref{MPIK}, M.~Lindner\thanksref{MPIK}, W.~Maneschg\thanksref{MPIK, Mirion}, M.~Meier\thanksref{MPIK}, S.~Mertens\thanksref{MPIK}, D.~Piani\thanksref{MPIK}, T.~Rink\thanksref{KIT}, H.~Strecker\thanksref{MPIK}
}

\authorrunning{E.~S\'{a}nchez Garc\'{i}a et al.}
\institute{Max-Planck-Institut f\"ur Kernphysik, Saupfercheckweg 1, 69117 Heidelberg, Germany \label{MPIK}  \and Institut f\"ur Astroteilchenphysik, Karlsruher Institut f\"ur Technologie (KIT), Hermann-von-Helmholtz-Platz 1, 76344 Eggenstein-Leopoldshafen, Germany\label{KIT} \and \emph{Present Address:} Marietta-Blau-Institut f\"{u}r Teilchenphysik der \"{O}AW, Dominikanerbastei 16, 1010 Wien, Austria\label{Wien} \and \emph{Present Address:} Mirion Technologies (Canberra) GmbH, Stahlstra\ss{}e 42--44, 65428 R\"usselsheim, Germany\label{Mirion}\vspace*{0.2cm} 
}
\thankstext{e1}{\href{mailto:esanchez@mpi-hd.mpg.de}{esanchez@mpi-hd.mpg.de (corresponding author)}}
\thankstext{e2}{\href{mailto:yulai.shi@mpi-hd.mpg.de}{yulai.shi@mpi-hd.mpg.de (corresponding author)}}
\thankstext{e3}{\href{mailto:conus.eb@mpi-hd.mpg.de}{conus.eb@mpi-hd.mpg.de}}

\date{\today}

\maketitle

\begin{abstract}
The \conusplus~experiment has recently reported the first detection of coherent elastic neutrino-nucleus scattering (\CEvNS) of reactor antineutrinos on germanium nuclei and is now entering a precision phase. The dominant uncertainty in the first measurement was the energy scale, which contributed 14\% to the uncertainty of the prediction of the combined signal. We present a dedicated neutron activation campaign in which one of the new 2.4~kg \conusplus~germanium detectors was irradiated with a strong $^{241}$AmBe source, demonstrating that a contribution below 4\% to the uncertainty of signal prediction is achievable. For the first time, the $^{71}$Ge \Mshell~X-ray line was clearly resolved at $(158.7\pm1.4)$~eV$_\text{ee}$, validating the \conusplus~energy reconstruction down to the detection threshold. This validation includes the understanding of the energy scale, the energy resolution, the trigger efficiency, and the correct separation of physical from noise events. These results establish the foundation for a future activation campaign at the Kernkraftwerk Leibstadt reactor site, strengthening the \conusplus~energy calibration and extending its sensitivity to precision \CEvNS~and beyond Standard Model physics measurements.
\end{abstract}

\keywords{Nucleus-neutrino interactions, Semiconductor detectors, High purity germanium detectors, Nuclear reactors, Low energy threshold}

\section{Introduction}

Coherent Elastic Neutrino-Nucleus Scattering (\CEvNS) is a Standard Model process in which neutrinos interact coherently with an entire nucleus via Z boson exchange. The scattering amplitudes from all nucleons add coherently, enhancing the cross-section by several orders of magnitude compared to other neutrino detection channels, such as inverse beta decay or elastic neutrino-electron scattering. Furthermore, the interaction is thresholdless because no charged leptons need to be produced. These properties make \CEvNS~a particularly appealing channel for reactor monitoring and for probing physics beyond the Standard Model (BSM)\cite{Lindner:2024eng,CentellesChulia:2025jir}. There is currently a broad worldwide effort to detect \CEvNS~at reactors~\cite{CONNIE:2024pwt, MINER:2025cevns, Colaresi:2022obx, NEON:2022hbk, NUCLEUS:2026pnv, nGeN:2025hsd, Ackermann:2024kxo, Akimov:2024lnl, Bailly-Salins:2026inm, Yang:2024exl, RELICS:2024opj, TEXONO:2024vfk}.

\CEvNS~was theoretically proposed in 1974~\cite{Freedman:1973yd,Kopeliovich:1974mv}, yet more than four 
decades elapsed before its first experimental detection by the COHERENT collaboration~\cite{Coherent:2017,COHERENT:2020iec,COHERENT:2026yje}, using neutrinos produced with a pion decay at rest source. The primary experimental challenge lies in the signature of the interaction: a nuclear recoil of only a few keV, requiring detectors with exceptionally low energy thresholds and backgrounds. More recently, \CEvNS~has been observed with $^{8}$B solar neutrinos~\cite{XENON:2026ydt,PandaX:2024muv, LZ:2025igz}, and with reactor antineutrinos by the \conusplus~experiment~\cite{Ackermann:2025obx}. A previous claim of reactor \CEvNS~detection~\cite{Colaresi:2022obx} remains in tension with the results of~\cite{Ackermann:2024kxo, TEXONO:2024vfk, nGeN:2025hsd}, underscoring the experimental difficulty of this measurement. Following the first \CEvNS~detection, \conusplus~now enters a precision phase. To this end, three of the original 1~kg high-purity germanium (HPGe) detectors have been replaced by new ones with 2.4~kg diodes, significantly increasing the target mass and thus the signal rate.

\begin{figure}
    \centering
    \includegraphics[width=0.48\textwidth]{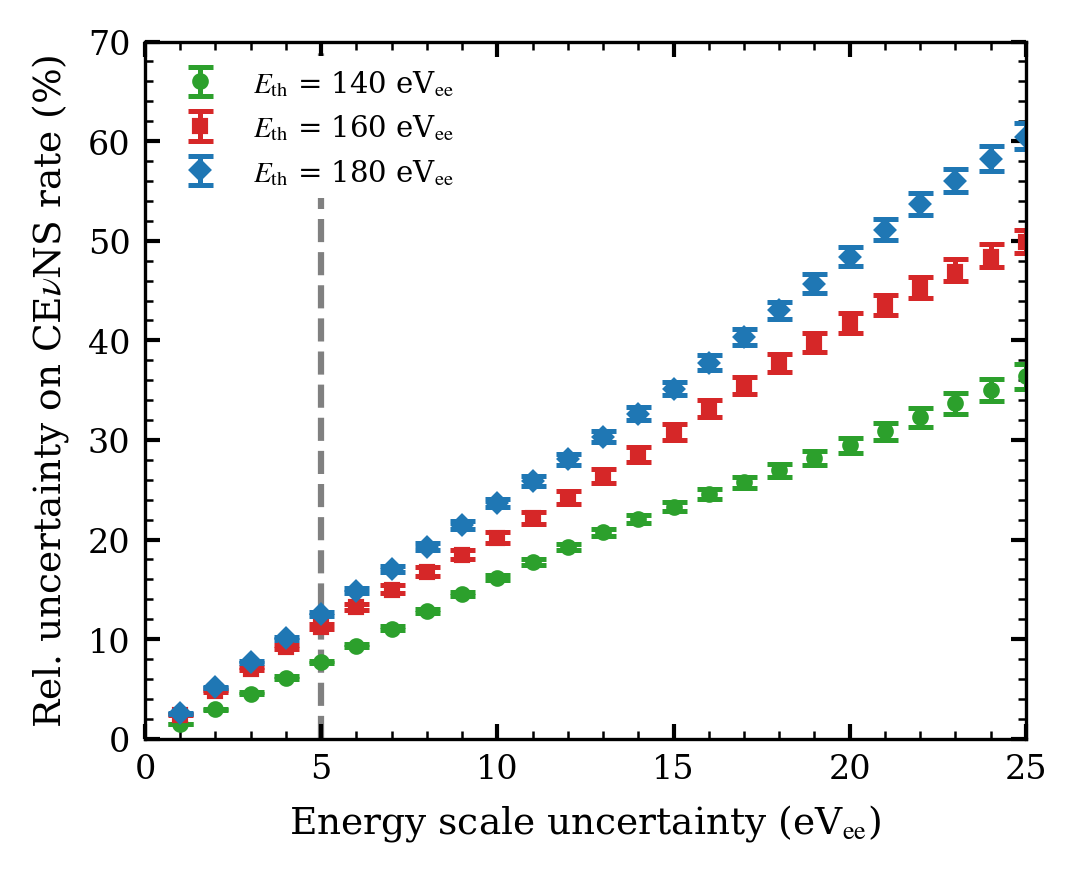}
    \caption{Signal prediction uncertainty as a function of the energy scale uncertainty, evaluated at a energy reference of 150~eV$_\text{ee}$ for three representative energy thresholds.}
    \label{unc_signal_prediction}
\end{figure}

The dominant uncertainty in the signal prediction in the first measurement was the energy scale, for which a conservative value of $\pm5$~eV$_\text{ee}$ (electron equivalent) was adopted. This resulted in an uncertainty of 14\% in the prediction of the combined signal, nearly twice the impact of the next largest contribution, the quenching factor, with an impact at the level of 7\%. Figure~\ref{unc_signal_prediction} shows the signal prediction uncertainty as a function of the energy scale uncertainty, evaluated at a reference energy of 150~eV$_\text{ee}$, for three realistic energy thresholds. The threshold dependence reflects the interplay between trigger efficiency and quenching. Reducing the energy scale uncertainty to $\pm3$~eV$_\text{ee}$ would render this contribution subdominant at the 4~\% level. Achieving a sub-percent impact on the signal prediction requires reaching an energy scale uncertainty below $\pm1$~eV$_\text{ee}$. 

External sources emitting $\gamma$-rays cannot be used for low energy calibration in \conusplus, as the emitted low energy $\gamma$-rays are absorbed in the end cap of the few-millimeter-thick copper cryostat and the dead layer of the detector. The calibration therefore relies on X-ray lines produced directly within the germanium crystal through cosmogenic and neutron-induced activation. During detector fabrication and transport above ground, cosmic-ray spallation produces $^{68}$Ge (half-life 270.95~days), $^{65}$Zn (half-life 244.01~days) and $^{68}$Ga (half-life 67.7~minutes). The latter is continuously produced by the decay of $^{68}$Ge. The $^{68}$Ge isotope is produced by spallation reactions on $^{70}$Ge, which require fast neutrons of at least $\sim$20~MeV~\cite{Barabanov:2006}:

\begin{equation}
    ^{70}\text{Ge} + n_\text{fast} \rightarrow \,^{68}\text{Ge} + 3n
    \label{eq:ge68_production}
\end{equation}

Similarly, $^{65}$Zn is produced by spallation reactions on the surface of the Earth~\cite{Barabanov:2006}:

\begin{equation}
    ^{70}\text{Ge} + n_\text{fast} \rightarrow \,^{65}\text{Zn} + 2n 
    + ^{4}\text{He}
    \label{eq:zn65_production}
\end{equation}

In addition to these cosmogenic isotopes, $^{71}$Ge is continuously produced \textit{in situ} within the detector via neutron capture on $^{70}$Ge (natural isotopic abundance 20.5\%) by thermal, epithermal and fast neutrons with energies up to a few MeV:

\begin{equation}
    ^{70}\text{Ge} + n \rightarrow \,^{71}\text{Ge} 
    \xrightarrow \,^{71}\text{Ga (stable)}
    \label{eq:ge71_production}
\end{equation}

Unlike $^{68}$Ge, which requires fast neutrons for its production through spallation and is therefore only activated above ground, $^{71}$Ge is produced continuously at the \conusplus~experimental site by muon-induced neutrons generated in the lead shielding, leading to a \Kshell~line count rate of $\sim$40~counts~kg$^{-1}$~day$^{-1}$~\cite{Ackermann:2025obx}. It can also be produced in larger quantities using external neutron sources such as $^{252}$Cf or $^{241}$AmBe.

Electron capture (EC) from all these isotopes produces characteristic X-ray lines via atomic de-excitation. The dominant feature in the low-energy range is the \Kshell~line at $10.367$~keV$_\text{ee}$, emitted with a branching ratio of $0.876$ from both $^{68}$Ge and $^{71}$Ge ($100\%$ EC). The corresponding \Lshell~lines appear at $1.298$~keV$_\text{ee}$ (L$_1$, branching ratio $\approx 0.119$) and $1.143$~keV$_\text{ee}$ (L$_2$, branching ratio $\approx 0.001$)~\cite{Agnese:2019cdmslite}. Due to the proximity in energy, the \Lshell~emissions from $^{68}$Ge, $^{71}$Ge, $^{68}$Ga and $^{65}$Zn are observed as a single broad peak even at the high energy resolution of the \conusplus\ detectors. Additionally, $^{68}$Ga contributes a \Kshell~line at $9.659$~keV$_\text{ee}$, at $11.1\%$ of the $^{68}$Ge \Kshell~intensity, and $^{65}$Zn produces a \Kshell~line at $8.979$~keV$_\text{ee}$, at $11.4\%$ of the $^{68}$Ge \Kshell~intensity. The \Mshell~line at $158.1$~eV$_\text{ee}$ is emitted by both $^{68}$Ge and $^{71}$Ge, but has not been resolved in previous CONUS campaigns due to insufficient statistics in the \Mshell~line. Contributions from copper activation in the cryostat are negligible. Since $^{68}$Ge and $^{71}$Ge produce X-ray lines at identical energies, their contributions are disentangled by exploiting the large difference in their half-lives ($270.95$~d vs.~$11.4645$~d) through a time-dependent fit of the \Kshell~line count rate~\cite{Bonet:2021wjw}.

The energy calibration of \conusplus~uses the \Kshell~(10.367~keV$_\text{ee}$) and \Lshell~(1.298~keV$_\text{ee}$) X-ray lines from $^{68}$Ge and $^{71}$Ge, with a linear interpolation assumed to extend into the \CEvNS~region of interest. The linearity of the data acquisition system and electronics has been studied via pulser scans~\cite{Ackermann:2025obx}, which previously revealed non-linearities at the level of up to 12~eV$_\text{ee}$ in the sub-keV regime. For the second phase, these non-linearities have been mitigated through the optimization of the energy reconstruction parameters, and alternative energy reconstruction algorithms are under development to further improve it.

\begin{figure}
    \centering
    \includegraphics[width=0.48\textwidth]{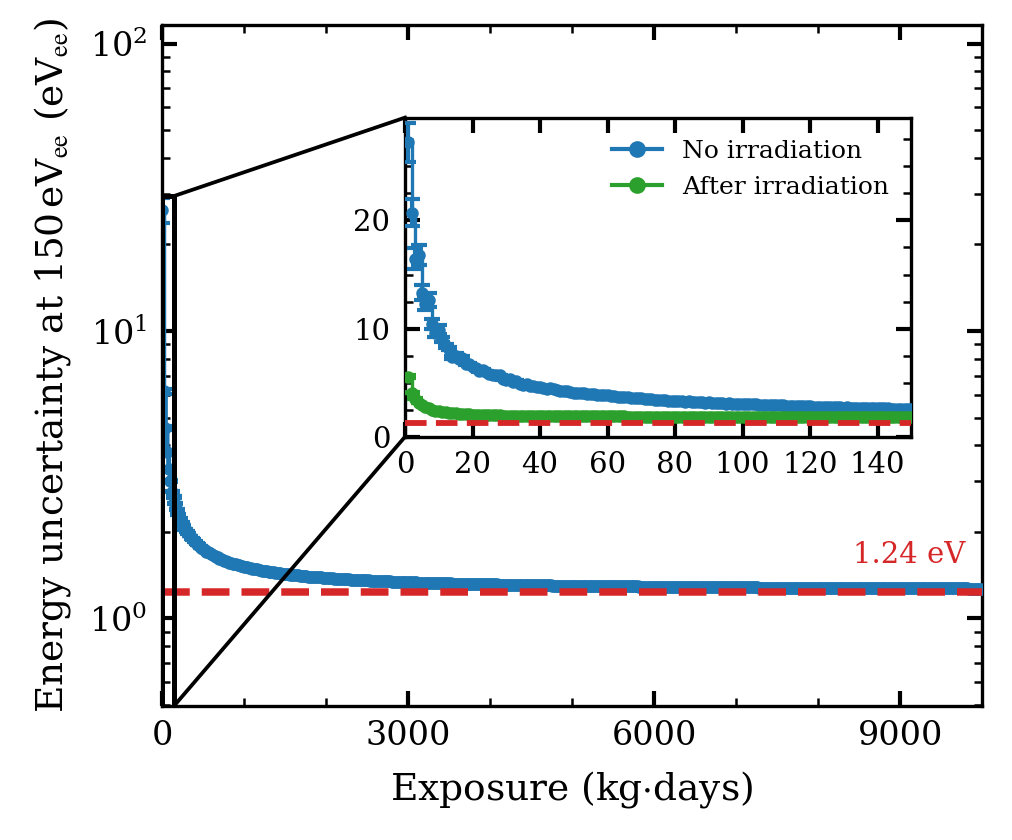}
    \caption{Energy scale uncertainty as a function of exposure for the nominal $^{71}$Ge production rate (blue) and an enhanced rate (green, $\times$10 factor). The dashed horizontal line indicates the target precision level, which is equivalent to the literature uncertainty on the peak positions~\cite{LNHB}.}
    \label{fig:unc_irradiation}
\end{figure}

Figure~\ref{fig:unc_irradiation} shows the energy scale uncertainty as a function of exposure, assuming a background rate consistent with \conusplus~operating conditions and a continuous $^{71}$Ge count rate of 40~counts~kg$^{-1}$~day$^{-1}$. Using the K- and \Lshell~lines, the precision floor is set by the uncertainty on the literature values of the peak positions~\cite{LNHB}, corresponding to a lower limit of approximately $\pm1.3$~eV$_\text{ee}$. Reaching this floor requires exposures of 4000~kg$\cdot$days, impractical given that the \conusplus~HPGe detectors have masses between 1 and 2.4~kg. Even achieving a $\pm3$~eV$_\text{ee}$ uncertainty requires more than 100~kg$\cdot$days, while sub-2~eV$_\text{ee}$ precision would demand exposures exceeding 400~kg$\cdot$days. The neutron activation of the detector with external sources allows to enhance the $^{71}$Ge production rate, in order to reduce the exposure needed. Assuming an enhancement factor of 10 (green line in Figure~\ref{fig:unc_irradiation}), an energy scale uncertainty below $\pm$2~eV$_\text{ee}$ can be achieved with only approximately 30~days of exposure, translating to a signal prediction uncertainty contribution below 4\%.

A further motivation for the neutron activation campaign is the enhancement of the \Mshell~X-ray line at 158.1~eV$_\text{ee}$. Observing this line serves multiple purposes: it directly validates the energy linearity of \conusplus~across the full \CEvNS~region of interest, provides an independent constraint on the trigger efficiency at threshold, yields an essential input for the background model, and can be used as a proxy for different noise and background rejection techniques as the pulse shape discrimination~\cite{Bonet:2023kob}.

In this work, we report the irradiation of one of the 2.4~kg \conusplus~HPGe detectors with an $^{241}$AmBe neutron source at the Max-Planck-Institut f\"ur Kernphysik (MPIK) in Heidelberg, Germany. We present the first clear observation of the $^{71}$Ge \Mshell~X-ray line, validating both the  activation procedure and the sub-keV detector response. This demonstration establishes the foundation for a future high-statistics activation campaign at the Kernkraftwerk Leibstadt (KKL) reactor site, which will significantly strengthen the \conusplus~energy calibration, extend the sensitivity to BSM physics in the \CEvNS~spectrum, and provide a robust validation of the trigger efficiency at the lowest detectable energies.

\section{Experimental setup}

The measurements were performed at the Low Level Laboratory (LLL) of the MPIK. The LLL provides an overburden of approximately 15~m of water equivalent (m~w.e.), consisting of rock, soil and concrete, which significantly suppresses the flux of cosmic-ray-induced neutrons and the hadronic component of cosmic rays~\cite{Heusser:1995wd}. The residual cosmic muon flux at this depth is $(65 \pm 2)$~muons~s$^{-1}$m$^{-2}$, corresponding to a reduction factor of $\sim$3 with respect to the surface~\cite{Heusser:1995wd}. The laboratory is equipped with an air conditioning system that regulates temperature and humidity and maintains $^{222}$Rn concentration in the range of $(30-70)$~Bq~m$^{-3}$. 

\begin{figure}
    \centering
    \includegraphics[width=0.48\textwidth]{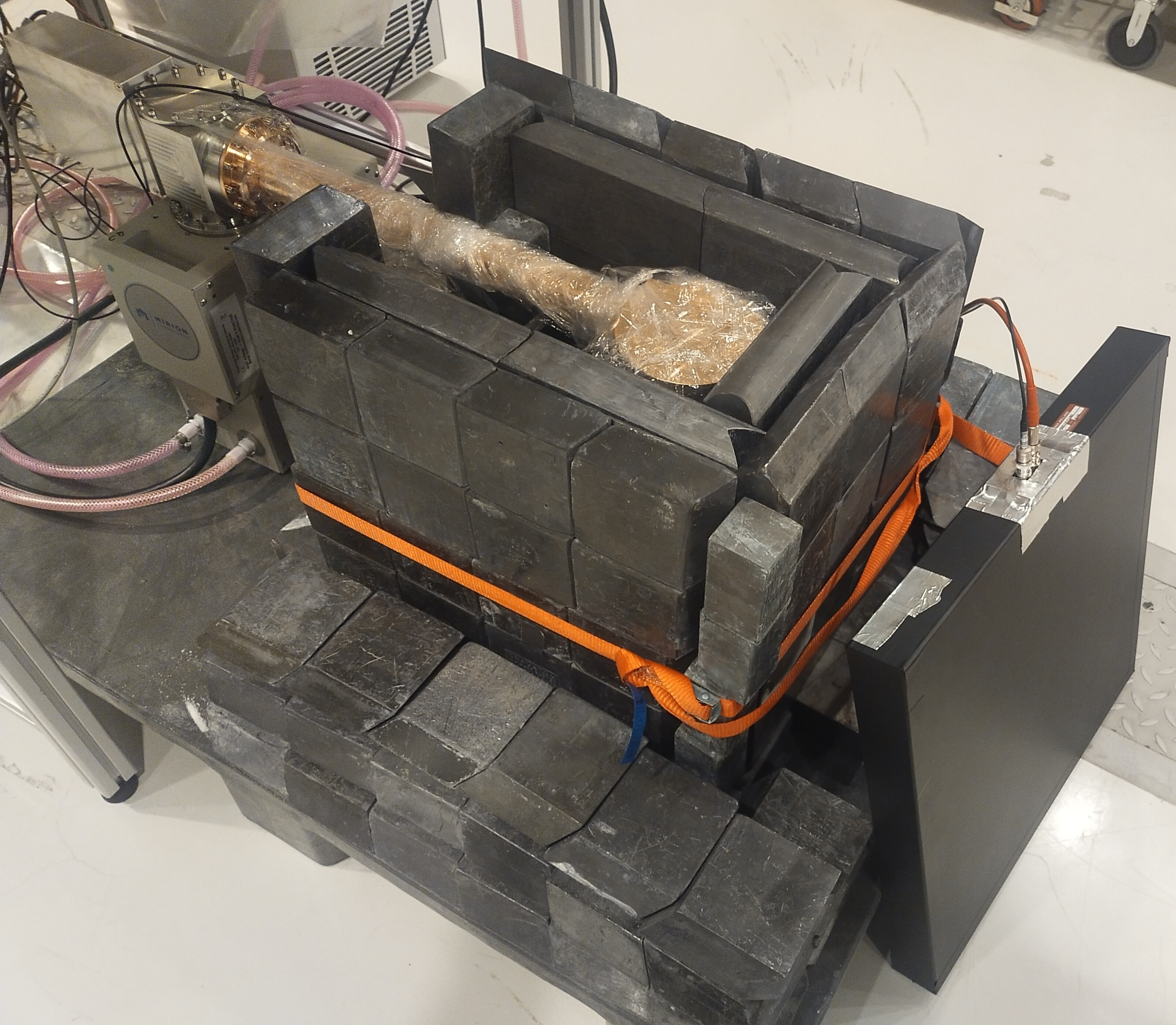}
    \caption{Photograph of the C8 detector setup at the MPIK Low Level Laboratory. The 2.4~kg HPGe detector is enclosed in a 10~cm lead shield. The upper part of the shield is open with the muon veto plate, which is on top during operation, in front of the shield.}
    \label{fig:setup}
\end{figure}

The experimental setup consists of a p-type point-contact (PPC) high-purity germanium (HPGe) detector, named C8, with a crystal mass of 2.4~kg, produced in the same batch as the new detectors installed at the \conusplus~experiment at KKL~\cite{CONUS:2024lnu}. The HPGe diode is cooled with a Cryo-Pulse 5 Plus (CP5+) electrically powered pulse tube cooler, upgraded with a liquid-cooled chiller system to suppress mechanical vibrations that would otherwise induce microphonic noise. The detector was operated under stable conditions, with the temperature maintained at $(20 \pm 1)^\circ$C by a dedicated air cooling system. The detector is equipped with a custom-built charge sensitive preamplifier (CSP) based on an application-specific integrated circuit (ASIC), providing improved noise performance and trigger efficiency at low energies. A pulsed-reset scheme generates rectangular inhibit signals (TRP) to veto spurious events after baseline resets. A full description of the detector and electronics upgrades is provided in~\cite{CONUS:2024lnu}. The detector is surrounded by 10~cm of lead on all sides to reduce the environmental $\gamma$-ray background. Additionally, an EJ-200 plastic scintillator 50$\times$50~cm$^{2}$ plate is placed on top of the shield and operated as an active muon veto. A photograph of the full setup is shown in Figure~\ref{fig:setup}. The data acquisition (DAQ) follows the same design as the \conusplus~experiment~\cite{CONUS:2024lnu}. It is based on a 16-bit CAEN V1782 digitizer sampling at 100~MHz managed via the CoMPASS software, with TRP inhibit signals recorded by a separate CAEN V1740D module. An external pulse generator (Keysight 33500B) injects artificial signals of known amplitude directly into the preamplifier, enabling regular monitoring of the energy resolution and trigger efficiency. The achieved pulser resolution in terms of full-width-at-half-maximum (FWHM) is (56~$\pm$~2)~eV$_\text{ee}$, and the trigger efficiency above 90\% is maintained down to 160~eV$_\text{ee}$.

Several data quality selection cuts are applied to the acquired data. A muon veto cut rejects all events within a 170~$\mu$s window following a muon trigger in the scintillator. A TRP cut vetoes events within 1000~$\mu$s after each preamplifier reset. A cut in the time difference distribution (TDD) rejects events separated by less than 1~ms from the preceding event, suppressing clustered microphonic noise in the low-energy region. Additionally, different quality cuts are applied based on the waveforms to remove pile-up and mis-reconstructed events. The combined dead time from all these contributions is accounted for in the final data analysis.

\section{Energy reconstruction}
\label{sec:energy_reco}

\begin{figure}
    \centering
    \includegraphics[width=0.48\textwidth]{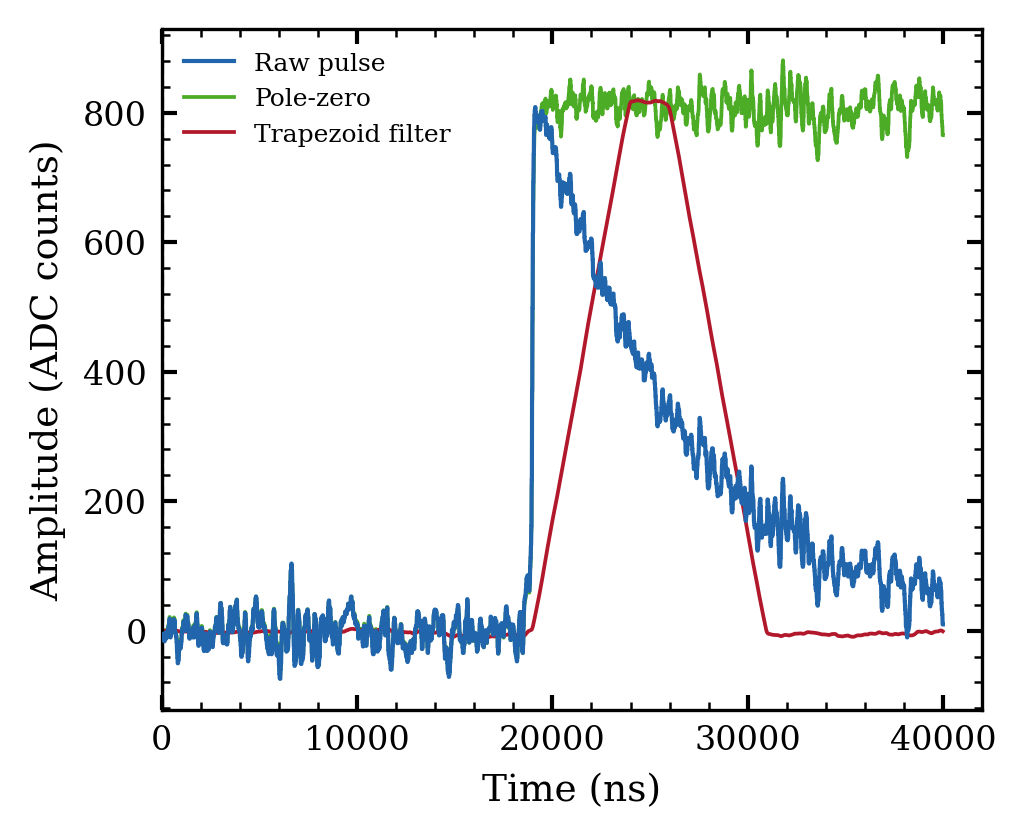}
    \caption{Representative detector pulse at three stages of the energy reconstruction chain: the raw preamplifier output (blue), the pole-zero corrected signal (green), and the trapezoidal filter output (red). The flat-top amplitude of the trapezoidal pulse is used as the energy estimator.}
    \label{fig:pulse}
\end{figure}

The signals produced by ionization events in the HPGe detector are processed through a chain of digital filtering steps to reconstruct the deposited energy. Figure~\ref{fig:pulse} shows a representative detector pulse at three stages of this chain. The raw preamplifier output (blue) exhibits an exponential decay tail produced by the AC coupling of the ADC input, with a decay time constant of 10~$\mu$s. A pole zero cancellation filter (green) is first applied to flatten this tail, yielding a step-like signal whose amplitude is proportional to the deposited energy. For each event, the baseline is estimated from the pre-trigger region of the waveform and subtracted prior to filtering, ensuring a stable reference level for the energy reconstruction. A trapezoidal filter (red) is then applied to the baseline-subtracted, pole-zero corrected signal, producing a shaped pulse whose flat-top amplitude is used as the energy estimator. The energy of each event is reconstructed online by the CAEN V1782 digitizer using this trapezoidal filter, implemented at the FPGA level via the CoMPASS software~\cite{CONUS:2024lnu}.

\begin{figure}
    \centering
    \includegraphics[width=0.48\textwidth]{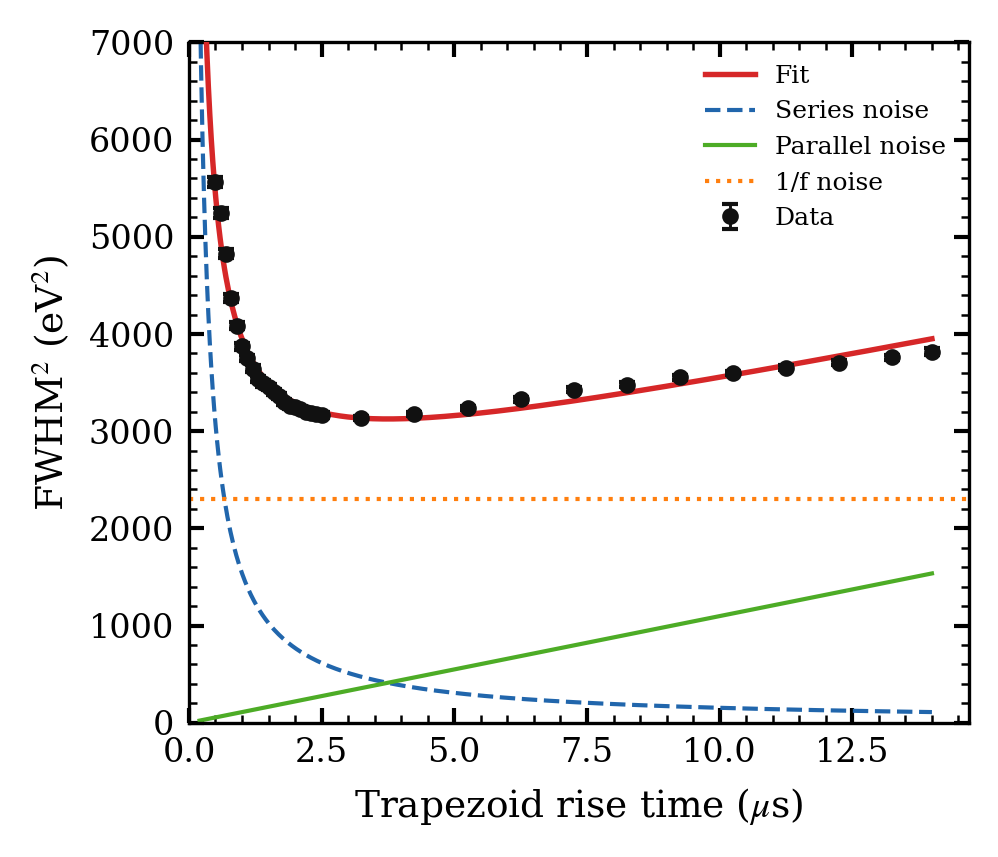}
    \caption{Squared energy resolution (FWHM$^2$) of the C8 detector as a function of the trapezoidal rise time, measured via pulser scans. The individual noise contributions are shown: $1/f$ noise (orange dotted), series noise (blue dashed), and parallel noise (green). The total fit (red) exhibits a minimum at $\sim$3.5~$\mu$s, which is adopted as the optimal working point.}
    \label{fig:noise}
\end{figure}

The choice of the trapezoidal rise time is a critical parameter that determines the noise performance of the energy reconstruction. Figure~\ref{fig:noise} shows the squared energy resolution (FWHM$^2$) as a function of the trapezoidal rise time, together with the individual noise contributions. Three components are identified: $1/f$ noise (orange), which is independent of the rise time and sets a constant floor; series noise (blue dashed), which dominates at short rise times and decreases rapidly with increasing shaping time; and parallel noise (green), which increases linearly with the rise time. The total FWHM$^2$ (red) exhibits a clear minimum at approximately 3~$\mu$s, where series and parallel noise are balanced. A trapezoidal rise time of 3.5~$\mu$s is therefore adopted for all data presented in this work, consistent with the value used for the \conusplus~detectors at KKL~\cite{CONUS:2024lnu}. For the flat top a value of  1~$\mu$s is considered.

The energy scale is established using the \Kshell~and \Lshell~X-ray lines from $^{68}$Ge and $^{71}$Ge, with a linear calibration function. However, the assumption of a purely linear DAQ response does not hold across the full energy range, in particular at sub-keV energies~\cite{Ackermann:2025obx}. The non-linearity is quantified via a pulser scan in the sub-keV range, defining the energy deviation as $\Delta E = E_\text{meas} - E_\text{real}$, where $E_\text{real}$ is derived from the known pulse amplitude and the linear calibration. The C8 detector exhibits a maximum deviation of $-12$~eV$_\text{ee}$ at $\sim$160~eV$_\text{ee}$, which is corrected through spline interpolation of the measured non-linearity. The impact of this correction on the \Mshell~line position is discussed in Section~\ref{sec:mshell}. For the \conusplus~Run-2 data collection at KKL, the CAEN DAQ settings were optimized, reducing the maximum non-linearity to $\sim$5~eV$_\text{ee}$ across the full sub-keV range.

A custom offline energy reconstruction algorithm has been developed for the \conusplus~experiment. By processing the full digitized waveform offline, this approach offers several advantages over the online CAEN reconstruction. Since the full waveform is available, advanced noise reduction techniques such as optimal filtering~\cite{Anderson:2022hgb} can be applied, which are not accessible in the online FPGA-based reconstruction. The non-linearity observed in the CAEN online reconstruction is suppressed in this approach, as the energy is extracted directly from the waveform without the constraints imposed by the fixed FPGA filter implementation. The custom reconstruction applied to the data demonstrates a comparable energy resolution to the online CAEN reconstruction, validating the approach. This algorithm will be used as the standard energy reconstruction for future \conusplus~analyses, providing a more accurate and stable energy scale at sub-keV energies, which is essential for high-precision \CEvNS~measurements.

\section{Neutron activation}
\label{sec:irradiation}

The detector was irradiated with an $^{241}$AmBe neutron source to enhance the $^{71}$Ge production rate within the crystal and improve the statistical precision of the low-energy calibration lines. The $^{241}$AmBe source produces neutrons through a two-step nuclear reaction process involving alpha decay from $^{241}$Am and subsequent interaction with beryllium nuclei:

\begin{equation}
\begin{aligned}
    ^{241}\text{Am} &\rightarrow \,^{237}\text{Np} + \alpha \\
    ^{9}\text{Be} + \alpha &\rightarrow \,^{12}\text{C} + n
\end{aligned}
\label{eq:ambe}
\end{equation}

Additional reaction channels, such as $^{9}$Be$(\alpha,n)^{8}$Be$+\alpha$, contribute at the few percent level. The source has an activity of 3.84~GBq corresponding to a neutron emission rate of approximately $10^5$~n/s. It emits neutrons with an average energy of about 4.5~MeV and a maximum energy up to 11~MeV~\cite{Lorch1973AmBe, Vijaya1973, Kluge1982}. This energy range is particularly well suited for the activation of $^{71}$Ge via neutron capture on $^{70}$Ge, while avoiding the production of long-lived radioactive isotopes in the surrounding detector materials, in particular copper. The activation products from copper, $^{64}$Cu (half-life 12.7~h) and $^{66}$Cu (half-life 5.12~min~\cite{ENSDF2022}), decay rapidly and their contributions are negligible for measurements performed several days after irradiation~\cite{NNDC:2025}. Similarly, the production of tritium via neutron-induced spallation of germanium requires neutron energies above $\sim$10~MeV~\cite{SuperCDMS:2018tqu} and is therefore strongly suppressed for the $^{241}$AmBe spectrum. Among all possible activation products, $^{71}$Ge is therefore the only isotope that produces a significant contribution. 

To maximize the neutron flux at the detector, the source-to-detector distance was reduced to 20~cm, compared to $\sim$1~m in the KKL irradiation campaigns, with fewer shielding layers between the source  and the detector. This configuration led to an estimated neutron flux of approximately 18~n~s$^{-1}$~cm$^{-2}$ at the detector position. The detector was continuously irradiated for one week, resulting in a total neutron fluence of approximately $1.1 \times 10^7$~n~cm$^{-2}$, well below the $10^9$~n~cm$^{-2}$ threshold for neutron-induced damage in germanium~\cite{Descovich2005NeutronDamage,Raudorf1987ChargeTrapping,Pehl1979RadiationDamage}. After irradiation, the detector was monitored for 31~days with daily spectra collected to study the time evolution of the activation products. Based on the detector mass of 2.4~kg, the $^{70}$Ge natural abundance and a neutron capture cross-section of 3~barn\cite{Mughabghab2006Atlas} for the $^{70}$Ge$(n,\gamma)^{71}$Ge reaction, approximately $10^8$ $^{71}$Ge atoms were produced during the irradiation. Taking into account the $^{71}$Ge half-life of (11.4645$\pm 0.0036)$~days~\cite{Derbin:2025}, this yields an expected emission of $\sim$$10^7$ K-shell,  $\sim$$10^6$ \Lshell~and $\sim$$10^5$ \Mshell~X-rays over the 31~days post-irradiation measurement period, representing an enhancement of more than three orders of magnitude compared to the $\sim$5000 \Kshell~counts previously observed in the KKL data. Prior to the irradiation, 40~days of background data were collected under identical conditions to enable a precise background subtraction. 

\begin{figure*}[ht]
    \centering
    \includegraphics[width=0.95\textwidth]{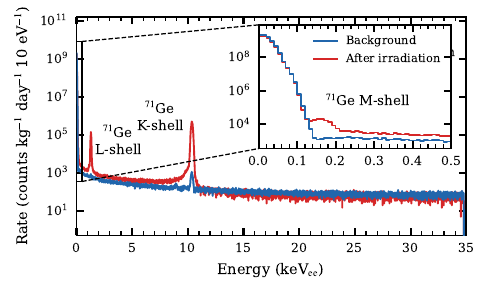}
    \caption{Comparison of the low-energy spectra acquired during 40~days before irradiation (blue) and 31~days after irradiation (red) with the $^{241}$AmBe neutron source. The $^{71}$Ge \Kshell~and \Lshell~X-ray lines become prominent after irradiation. The inset shows the sub-500~eV$_\text{ee}$ region, where an excess consistent with the $^{71}$Ge \Mshell~line at 158.1~eV$_\text{ee}$ is visible after irradiation.}
    \label{fig:spectra_irradiation}
\end{figure*}

Figure~\ref{fig:spectra_irradiation} shows the comparison of the low-energy spectra up to 35~keV$_\text{ee}$ acquired before (blue) and after (red) irradiation. The first 2 days after irradiation were not included to remove the short half-life products from copper.  After irradiation, the $^{71}$Ge \Kshell~line at 10.367~keV$_\text{ee}$ and \Lshell~line at 1.298~keV$_\text{ee}$ become prominent, with count rates enhanced by more than three orders of magnitude with respect to the pre-irradiation background. The inset shows the region below 500~eV$_\text{ee}$, where a clear excess above the background is visible after irradiation, consistent with the expected \Mshell~X-ray emission from $^{71}$Ge at 158.1~eV$_\text{ee}$. The analysis of the \Mshell~line is presented in detail in Section~\ref{sec:mshell}.

\begin{figure*}[ht]
    \centering
    \includegraphics[width=0.48\textwidth]{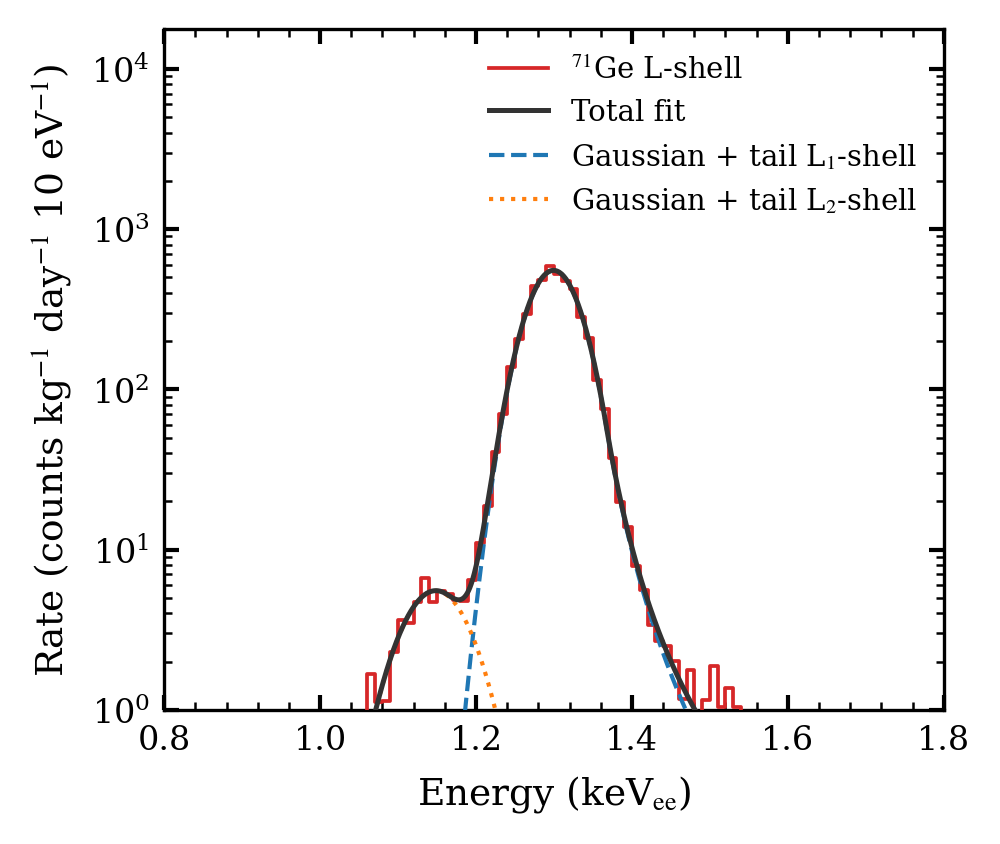}
    \hfill
    \includegraphics[width=0.48\textwidth]{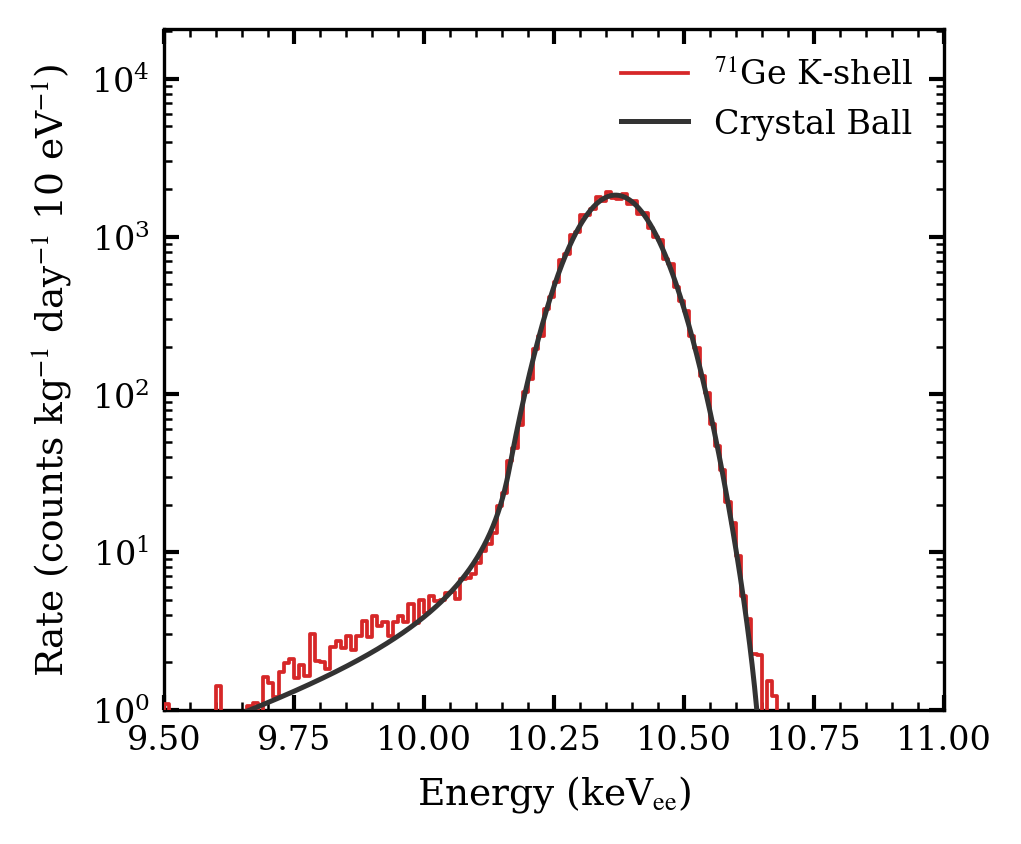}
    \caption{Fits to the $^{71}$Ge \Lshell~(left) and \Kshell~(right) X-ray lines after subtracting the background before irradiation. The \Lshell~data (red) are fitted with two Gaussian functions (black) accounting for the L$_1$ ($1298.5$~eV$_\text{ee}$) and L$_2$ ($1148.2$~eV$_\text{ee}$) sub-shell capture lines. The \Kshell~data (red) are fitted with a Crystal Ball function (black), which accounts for the asymmetric low-energy tail arising from incomplete charge collection.}
    \label{fig:kl_shells}
\end{figure*}

The K- and \Lshell~lines are used to establish the low-energy calibration of the C8 detector after irradiation. The spectra shown in Figure~\ref{fig:kl_shells} are obtained after subtraction of the background before irradiation, isolating the contribution from the $^{71}$Ge decay.  The \Lshell~region (left panel) is fitted with two Gaussian functions with a high energy tail to account for the L$_1$ and L$_2$ sub-shell capture lines, with fitted centroids of $(1298.5 \pm 1.3)$~eV$_\text{ee}$ and $(1148.2 \pm 6.1)$~eV$_\text{ee}$~\cite{xdb_table1,Agnese:2019cdmslite}, respectively, in good agreement with the reference values of $1298$~eV$_\text{ee}$ and $1143.2$~eV$_\text{ee}$. The asymmetric high-energy tail in both lines is attributed to pile-up events. The \Kshell~(right panel) peak at $10367$~eV$_\text{ee}$ is fitted with a Crystal Ball function, which accounts for the asymmetric low-energy tail arising from incomplete charge collection in the detector bulk and surface regions. As a cross-check, the fits are repeated without the background subtraction. In this case, a step-like background component is included for the \Kshell~region, while an exponential background is assumed for the L-shell. The resulting fit parameters are consistent within 1\%. The total number of counts in the \Kshell~peak is of the order of $10^7$, consistent with the expectations from the activation calculation described above. The calibration achieved after irradiation yields an energy scale uncertainty of $\pm1.3$~eV$_\text{ee}$ at 150~eV$_\text{ee}$, a significant improvement with respect to the pre-irradiation value of more than $\pm5$~eV$_\text{ee}$. For each line, the reconstructed peak position, energy resolution (FWHM), total count rate, and the ratio of the \Kshell~to \Lshell~count rates are summarized in Table~\ref{tab:shells}, together with the comparison to the literature values~\cite{LNHB}.

To confirm that the observed lines originate from $^{71}$Ge decay, the time evolution of the K- and \Lshell~peak count rates was studied using daily spectra collected over the 31-days after the irradiation period. Both decay curves are well described by a single exponential function, yielding half-life values of $(11.46 \pm 0.02)$~days for the \Kshell~and $(11.49 \pm 0.16)$~days for the L-shell. These values are in excellent agreement with the most precise recent measurement of the $^{71}$Ge half-life, $T_{1/2} = (11.4645 \pm 0.0036)$~days~\cite{Derbin:2025}, and with the other recent measurements~\cite{Collar:2023, Norman:2024}. The precise knowledge of the $^{71}$Ge half-life is of broader interest beyond detector calibration, as it plays a direct role in the interpretation of the so-called gallium anomaly discussed in the context of the SAGE, GALLEX/GNO and BEST~\cite{SAGE:2009,Kaether:2010,BEST:2022} experiments, which was proposed to be partially explained by an underestimation of the $^{71}$Ge half-life~\cite{Giunti:2023}. The recent precise measurements exclude this hypothesis. Our independent measurement, obtained as a byproduct of the calibration campaign, is fully consistent with these results and provides an additional cross-check of the $^{71}$Ge half-life in a completely different experimental context. The large statistics accumulated after irradiation also enable, for the first time in \conusplus, the study of the $^{71}$Ge \Mshell~X-ray line at $\sim$158.1~eV$_\text{ee}$, which was not accessible in previous campaigns due to the insufficient activation rate. The analysis of this line is presented in the following section.

\section{\Mshell~line observation}
\label{sec:mshell}

The $^{71}$Ge \Mshell~X-ray line at $\sim$158.1~eV$_\text{ee}$ lies in the most challenging region of the detector response, close to the energy threshold just above the region dominated by electronic noise. Its observation requires both a sufficient activation rate, as achieved by the $^{241}$AmBe irradiation campaign described in Section~\ref{sec:irradiation}, and a precise energy reconstruction. 

\begin{figure*}[ht]
    \centering
    \includegraphics[width=0.48\textwidth]{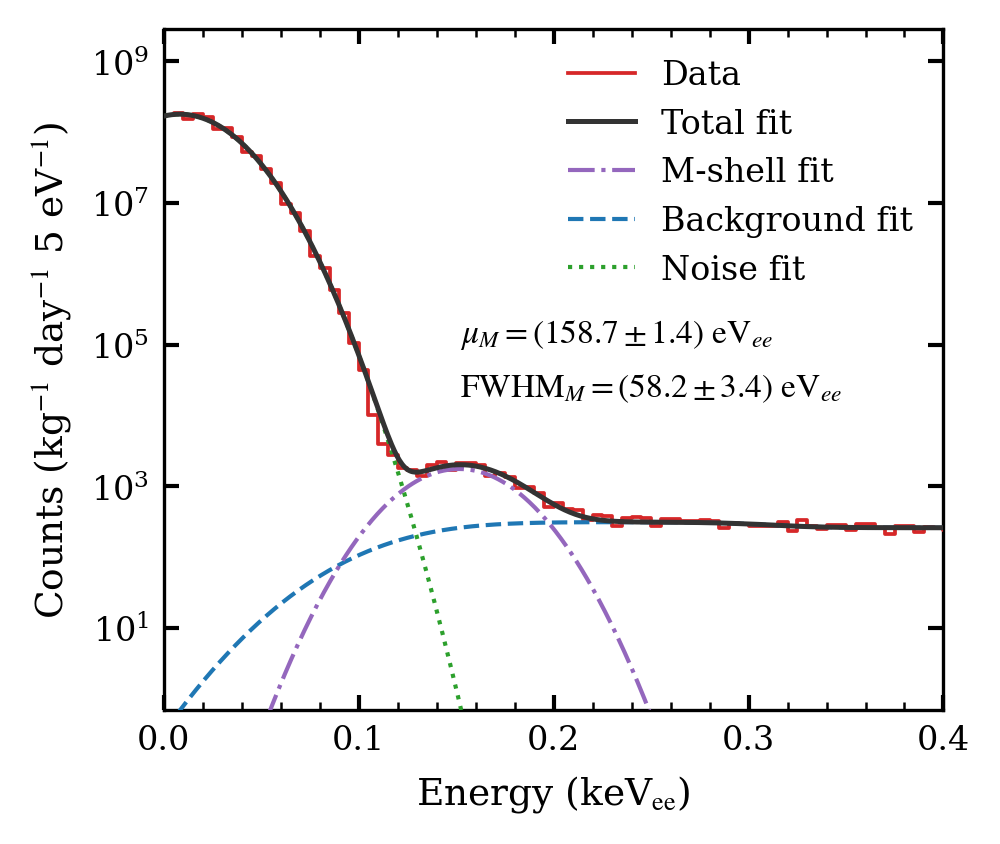}
    \hfill
    \includegraphics[width=0.48\textwidth]{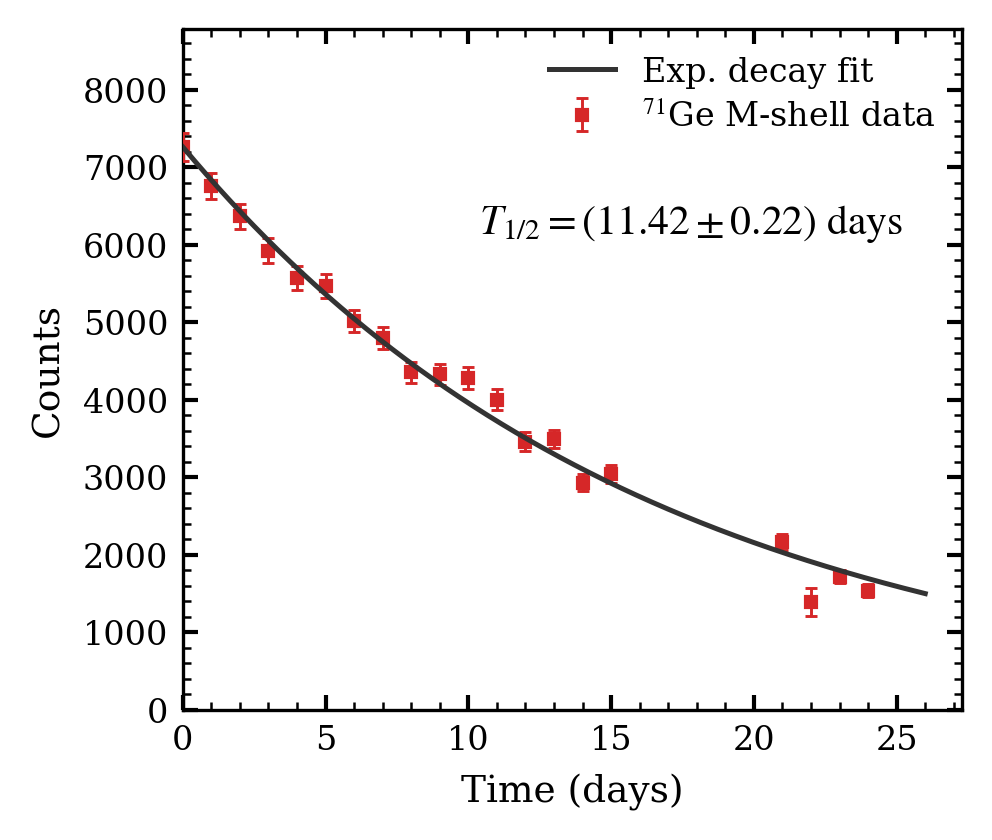}
    \caption{Left: fit of the $^{71}$Ge \Mshell~X-ray line in the  post-irradiation spectrum. The data (red) are described by a model consisting of a noise pre-fit (green), an exponential background weighted by the trigger efficiency (blue dashed), and the convolution of the \Mshell~signal with the trigger efficiency and electronic smearing (black), yielding $\mu_M = (158.7\pm1.4)$~eV$_\text{ee}$ and FWHM$_M = (58.2\pm3.4)$~eV$_\text{ee}$. Right: time evolution of the daily \Mshell~yield $N_M$ over the 31-day post-irradiation period (blue points), fitted with a single exponential decay function (red), yielding a half-life of $(11.42 \pm 0.22)$~days, consistent with the known $^{71}$Ge half-life of $(11.4645 \pm 0.0036)$~days~\cite{Derbin:2025}.}
    \label{fig:mshell}
\end{figure*}

The \Mshell~peak is extracted from the post-irradiation spectrum without subtraction of the background to avoid effects related to the noise peak stability. The resulting spectrum in the $[0, 400]$~eV$_\text{ee}$ range is shown in the left panel of Figure~\ref{fig:mshell}. The spectrum is dominated at low energies by the noise peak, which is pre-fitted independently ($f_\text{noise}(E)$). The remaining spectrum is described by a fit model consisting of the trigger efficiency $\varepsilon_\text{trig}$ multiplied by the sum of an exponential background and a Gaussian peak for the \Mshell~line:

\begin{multline}
    f(E) = f_\text{noise}(E) + \varepsilon_\text{trig}(E) \cdot e^{a - bE} \\
    + A \left[ \varepsilon_\text{trig} \cdot 
    \mathcal{G}(\mu_M, \sigma_\text{Fano}) \right] \ast 
    \mathcal{G}(0, \sigma_\text{noise}),
    \label{eq:mshell_fit}
\end{multline}

where $a$ and $b$ are the exponential background parameters, $\mu_M$ is the \Mshell~peak position, $A$ the amplitude, $\sigma_\text{noise}$ the contribution of the electronic noise to the \Mshell~width and $\sigma_\text{Fano}$ the contribution from the Fano factor. The trigger efficiency $\varepsilon_\text{trig}$ acts on the true energy $E'$ before the electronic smearing, as there is no correlation between the DAQ trigger filter and the trapezoidal energy reconstruction. Since $\sigma_\text{Fano} \ll \sigma_\text{noise}$, the peak shape is nearly symmetric. The total energy resolution (FWHM$_M$) is obtained by combining the two contributions in quadrature. The fit yields a peak position of $\mu_M = (158.7\pm1.4)$~eV$_\text{ee}$ and an energy resolution of  FWHM$_M = (58.2\pm3.4)$~eV$_\text{ee}$, in good agreement with the literature value of $(158.1\pm0.5)$~eV$_\text{ee}$~\cite{LNHB}. The stability of the fit and the uniqueness of the minimum were verified through a 2D $\Delta\chi^2$ profile scan in the ($\mu_M$, FWHM$_M$) plane, confirming well-defined confidence contours. The small correlation between the peak position and width is accounted for in the propagation of the energy calibration systematic into the resolution uncertainty.

To confirm the $^{71}$Ge origin of the observed \Mshell~peak, the time evolution of the \Mshell~yield was studied using daily spectra after irradiation, as shown in the right panel of Figure~\ref{fig:mshell}. The decay curve is well described by a single exponential function, yielding a half-life of ($11.42 \pm 0.22)$~days, in good agreement with the known $^{71}$Ge half-life of $(11.4645 \pm 0.0036)$~days~\cite{Derbin:2025} and consistent with the values obtained for the K- and \Lshell~lines, confirming that the observed peak originates from $^{71}$Ge decay.

The reconstructed peak positions and energy resolutions for the K-, L- and \Mshell~lines are summarized in Table~\ref{tab:shells}, together with the comparison to the literature values~\cite{LNHB}. The measured positions are in good agreement for all three shells, validating the energy calibration, trigger efficiency and the non-linearity correction across the full sub-keV range from 158.1~eV$_\text{ee}$ to 10.367~keV$_\text{ee}$. The K/L and M/L ratios of the count rates are also consistent with the expected branching ratios~\cite{LNHB}, indicating a proper modeling of the trigger efficiency. As an additional cross-check, the fit was repeated leaving the trigger efficiency parameters free, yielding values consistent with the independently measured ones, further validating the event reconstruction at sub-keV energies.

\begin{table*}[ht]
\caption{Summary of the fit results for the $^{71}$Ge K-, L- and \Mshell~
X-ray lines, compared to literature values~\cite{LNHB, Derbin:2025, 
Agnese:2019cdmslite}. The reconstructed peak position, energy resolution 
(FWHM), and intensity ratios with respect to the \Kshell~are reported.}
\label{tab:shells}
\centering
\setlength\extrarowheight{3pt}
\resizebox{\textwidth}{!}{%
\begin{tabular}{llcccc}
\hline
& & K-shell & L$_1$-shell & L$_2$-shell & M-shell \\
\hline
\multirow{2}{*}{Position [eV$_\text{ee}$]}
    & This work  & $10368.3\pm1.3$ & $1298.5\pm1.3$ & $1148.2\pm6.1$ & $158.7\pm1.4$ \\
    & Literature & $10367.1\pm0.5$ & $1297.7\pm1.1$ & $1143.2$       & $158.1\pm0.5$ \\
\hline
{FWHM [eV$_\text{ee}$]}
    & This work  & $170.6\pm1.3$ & $76.4\pm1.3$ & $73.8\pm7.4$ & $58.2\pm3.4$ \\
\hline
\multirow{2}{*}{Ratio to K-shell}
    & This work  & $1$ & $0.118\pm0.001$ & $0.0014\pm0.0003$ & $ 0.0214\pm0.0004$ \\
    & Literature & $1$ & $0.119\pm0.003$ & $0.001$ & $0.0205$           \\
\hline
\multirow{2}{*}{Half-life [days]}
    & This work  & $11.46\pm0.02$ & \multicolumn{2}{c}{$11.49\pm0.16$} & $11.42\pm0.22$ \\
    & Literature & \multicolumn{4}{c}{$11.4645\pm0.0036$} \\
\hline
\end{tabular}%
}
\end{table*}

\begin{figure}
    \centering
    \includegraphics[width=0.48\textwidth]{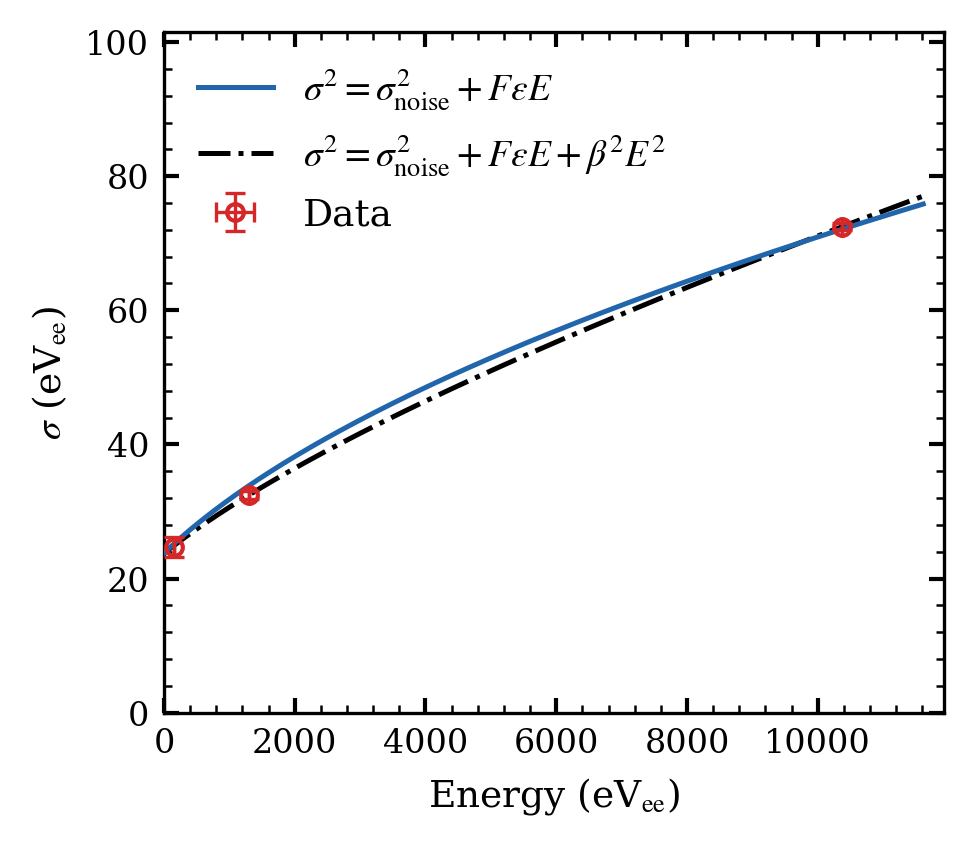}
    \caption{Energy resolution ($\sigma$) as a function of energy measured at the $^{71}$Ge M-, L- and \Kshell~X-ray lines at 158.7, 1298.5 and 10368.3~eV$_\text{ee}$, respectively. The solid blue line shows a fit of the standard Fano model ($F = 0.151 \pm 0.002$),  and the black dash-dotted line includes the quadratic term for incomplete charge collection ($F = 0.123 \pm 0.011$). In both cases $\sigma_\text{noise}$ is fixed by the pulser measurement (FWHM ($56 \pm 2$)~eV$_\text{ee}$). The Fano factor including the quadratic term is in better agreement with the values in literature~\cite{Croft:1996, Lowe:1997, Papp2005Fano, Wiens2013Germanium71, Salathe2015Fano, Agostini2019GERDA}.} 
    \label{fig:resolution}
\end{figure}

The energy resolution of ($58.2 \pm 3.4$)~eV$_\text{ee}$ (FWHM) measured at the \Mshell~line at ($158.7 \pm 1.4$)~eV$_\text{ee}$ provides a direct experimental determination of the detector resolution at sub-keV energies. The energy resolution of a Ge detector is dominated by two main contributions, the electronic noise and the Fano fluctuations in the charge creation process. Other sources like microphonic noise are found to be negligible. Effects such as incomplete charge collection can be accounted for with a quadratic term at high energies~\cite{Lepy:2000, SuperCDMS:2018gro}. The total resolution is described by:
\begin{equation}
    \sigma^2(E) = \sigma_\text{noise}^2 + F \cdot \epsilon \cdot E + \beta^2 \cdot E^2,
    \label{eq:fwhm_total}
\end{equation}
where $F$ is the dimensionless Fano factor, $\epsilon = 2.96$~eV is the mean energy required to create an electron-hole pair in germanium~\cite{Lepy:2000}, and $\beta$ is the dimensionless ballistic deficit parameter accounting for incomplete charge collection at higher energies~\cite{SuperCDMS:2018gro}. In the absence of ballistic deficit ($\beta = 0$), Eq.~\ref{eq:fwhm_total} reduces to the standard Fano model. The electronic noise contribution $\sigma_\text{noise}$ is fixed from the pulser measurement (($56 \pm 2$)~eV$_\text{ee}$ FWHM). The total energy resolution as a function of energy is shown in Figure~\ref{fig:resolution}. Fitting the standard Fano model ($\beta = 0$, blue solid line) to the M-, L-and \Kshell~data yields $F = 0.151 \pm 0.002$. Including the quadratic term (black dash-dotted line) significantly improves the agreement with the data, giving $F = 0.123 \pm 0.011$ and a ballistic deficit parameter of $\beta = (2.9 \pm 0.5) \times 10^{-3}$.

The Fano factor from the quadratic fit is in good agreement with values in the range $0.08-0.13$ typically reported in the literature for HPGe detectors~\cite{Croft:1996, Lowe:1997, Papp2005Fano, Wiens2013Germanium71, Salathe2015Fano, Agostini2019GERDA}. It is also in good agreement with the value used in~\cite{Ackermann:2025obx}. The value is significantly smaller than $F = 0.39$ reported by SuperCDMS~\cite{SuperCDMS:2018gro}, which may suggest a temperature dependence of the Fano factor, given that the present detector was operated at 73~K while SuperCDMS operates at $\sim$50~mK. The ballistic deficit term is similar in both detectors. 

\section{Conclusions}
\label{sec:conclusions}

In this work, we have presented a dedicated neutron activation campaign aimed at improving the sub-keV energy calibration of the \conusplus~experiment and demonstrated the detector response at energies close to the detection threshold. The campaign was performed at the Low Level Laboratory of MPIK using the C8 detector, one of the new 2.4~kg HPGe detectors produced for the \conusplus~Phase-2, and a $^{241}$AmBe neutron source placed at 20~cm from the detector. The resulting neutron fluence of $\sim$$1.1 \times 10^7$~n~cm$^{-2}$ produced approximately $10^8$ $^{71}$Ge atoms within the crystal, enhancing the K- and \Lshell~X-ray line rates by more than three orders of magnitude compared to the previous irradiation at KKL. The use of neutron-induced $^{71}$Ge activation provides several key advantages for low-energy calibration. Its relatively short half-life of about 11 days allows the detector to return to nominal background conditions within a few months, enabling repeated calibrations between physics runs. Additionally, these events originate mostly from bulk energy depositions, closely matching the topology of \CEvNS~interactions and therefore providing an optimal proxy for signal-like events.

The enhanced $^{71}$Ge statistics allowed a significant improvement of the energy scale calibration. The energy scale uncertainty at 150~eV$_\text{ee}$, the reference threshold for the \CEvNS~analysis, was reduced from more than $\pm5$~eV$_\text{ee}$ before irradiation to $\pm1.3$~eV$_\text{ee}$ after irradiation. Propagating this improvement into the \conusplus~signal prediction, the contribution of the energy scale uncertainty to the combined signal prediction uncertainty is reduced from 14\% to below 4\%, making it comparable to other uncertainties. This demonstrates that a single activation campaign of one week is sufficient to reach the calibration precision required for the upcoming \conusplus~phases.

For the first time, the $^{71}$Ge \Mshell~X-ray line at $\sim$158~eV$_\text{ee}$ has been clearly resolved in a \conusplus~ detector. The measured peak position of ($158.7\pm1.4$)~eV$_\text{ee}$ is in good agreement with the literature value, and the energy resolution (FWHM) of ($58.2\pm3.4$)~eV$_\text{ee}$ is consistent with the expected detector performance at this energy. The $^{71}$Ge origin of the line was confirmed by the time evolution of the \Mshell~yield, which follows an exponential decay with a half-life of ($11.42\pm 0.22$)~days, in good agreement with the known value of ($11.4645\pm0.0036$)~days~\cite{Derbin:2025}. The \Mshell~signal is observed with a statistical significance greater than $10\sigma$. Beyond the validation of the energy scale, the clear observation of the \Mshell~line demonstrates the detector performance in the immediate vicinity of the \CEvNS~region of interest, providing experimental access to the energy resolution, the trigger efficiency turn-on, and the capability to discriminate physical low-energy depositions from noise and background events at threshold.

The energy resolution measured at the M-, L- and \Kshell~energies is well described by the resolution model of Eq.~\ref{eq:fwhm_total}. A Fano factor of $F = 0.123 \pm 0.011$ is estimated consistent with values reported in the literature~\cite{Croft:1996, Lowe:1997, Papp2005Fano, Wiens2013Germanium71, Salathe2015Fano, Agostini2019GERDA}. The non-linearity of the CAEN online energy reconstruction was characterized via dedicated pulser scans, revealing a maximum deviation of $-12$~eV$_\text{ee}$ at $\sim$160~eV$_\text{ee}$. A spline interpolation correction was applied to all reconstructed energies prior to the spectral analysis, validated
by the agreement of the measured \Mshell~peak position with the literature value.

The results presented in this work establish the full experimental and analysis foundation for a future high-statistics activation campaign at the KKL reactor site. Applying the same irradiation procedure with the $^{252}$Cf source available at KKL will allow the \conusplus~detectors to achieve an energy scale uncertainty below 2~eV$_\text{ee}$, reducing the signal prediction uncertainty contribution to below 4\% for all detectors simultaneously. Combined with the validated trigger efficiency and the demonstrated \Mshell~response, this will significantly strengthen the \conusplus~energy calibration and open the path to precision measurements of the \CEvNS~cross-section in the fully coherent regime. In addition, the high-statistics sample of $^{71}$Ge decays, dominated by bulk events, provides an excellent dataset to further develop pulse shape discrimination techniques, enabling improved separation of surface and bulk events in future high-precision \CEvNS~measurements.

\small

\textbf{Acknowledgements}
We thank Mirion Technologies (Canberra) in Lingolsheim for their highly professional support. 

\normalsize


\bibliographystyle{bibliostyle}
\bibliography{references}
\end{document}